\documentclass[twocolumn,english,reprint,aip,citeautoscript]{revtex4-1}
\usepackage[LGR,T1]{fontenc}
\usepackage[latin9]{inputenc}
\usepackage{verbatim}
\usepackage{amsmath}
\usepackage{graphicx}
\usepackage{subscript}
\usepackage{placeins}

\makeatletter

\DeclareRobustCommand{\greektext}{%
  \fontencoding{LGR}\selectfont\def\encodingdefault{LGR}}
\DeclareRobustCommand{\textgreek}[1]{\leavevmode{\greektext #1}}
\ProvideTextCommand{\~}{LGR}[1]{\char126#1}

\usepackage{algorithm,algpseudocode}
\makeatletter
\newcommand{\algmargin}{\the\ALG@thistlm}
\makeatother
\newlength{\whilewidth}
\algdef{SE}[parWHILE]{parWhile}{EndparWhile}[1]
  {\parbox[t]{\dimexpr\linewidth-\algmargin}{%
     \hangindent\whilewidth\strut\algorithmicwhile\ #1\ \algorithmicdo\strut}}{\algorithmicend\ \algorithmicwhile}%
\algnewcommand{\parState}[1]{\State%
  \parbox[t]{\dimexpr\linewidth-\algmargin}{\strut #1\strut}}

\makeatother

\usepackage{babel}
\begin{document}

\title{Even-handed subsystem selection in projection-based embedding}

\author{Matthew Welborn}

\affiliation{Division of Chemistry and Chemical Engineering, California Institute
of Technology, Pasadena, CA 91125}

\author{Frederick R. Manby}

\affiliation{Centre for Computational Chemistry, School of Chemistry, University
of Bristol, Bristol BS8 1TS, United Kingdom}

\author{Thomas F. Miller III}

\affiliation{Division of Chemistry and Chemical Engineering, California Institute
of Technology, Pasadena, CA 91125}
\begin{abstract}
Projection-based embedding offers a simple framework for embedding
correlated wavefunction methods in density functional theory. %
Partitioning between the correlated wavefunction and density functional
subsystems is performed in the space of localized molecular orbitals.
However, during a large geometry change \textendash{} such as a chemical
reaction \textendash{} the nature of these localized molecular orbitals,
as well as their partitioning into the two subsystems, can change
dramatically. This can lead to unphysical cusps and even discontinuities
in the potential energy surface. In this work, we present an even-handed
framework for localized orbital partitioning that ensures consistent
subsystems across a set of molecular geometries. We illustrate this
problem and the even-handed solution with a simple example of an S\textsubscript{N}2
reaction. Applications to a nitrogen umbrella flip in
a cobalt-based CO\textsubscript{2} reduction catalyst and to the
binding of CO to Cu clusters are presented. In both cases, we find that even-handed
partitioning enables chemically accurate embedding with modestly-sized
embedded regions for systems in which previous partitioning strategies are problematic.
\end{abstract}
\maketitle

\section{Introduction}

Embedding methods for electronic structure offer the possibility of accurate description of chemical
reactions at greatly reduced computational cost by treating different
parts of a chemical system with different levels of theory. Many versions of
embedding exist, including implicit solvent,\cite{Born1920,Miertus1981,Cramer1999}
QM/MM~\cite{Warshel1976b,Field1990,Lin2006} and ONIOM,\cite{Vreven2006}
subsystem density functional theory
(DFT) and frozen density embedding,\cite{Cortona1991,Wesolowski1993,Wesolowski1994,Iannuzzi2006,Roncero2008,Elliott2010,Fux2010,Huang2011,Nafziger2011,Goodpaster2010,Goodpaster2011,Goodpaster2012,Manby2012,SeveroPereiraGomes2012,Goodpaster2014a,Libisch2014,Wesolowski2015}
active space methods,\cite{Roos1980,Malmqvist1990} local correlation
treatments,\cite{Pulay1983,Werner2006} Green's function embedding,\cite{Inglesfield1981,Zgid2010,Phillips2014}
and density matrix embedding theory.\cite{Knizia2012,Knizia2013}
Central to all of these methods is the question of how the system is
to be partitioned into a number of subsystems to be treated at different
levels of theory. 

Prescriptions for partitioning the system usually consider a single
geometry at a time. However, this approach may fail when considering a
chemical reaction in which the nature of the embedded subsystem changes
across the reaction coordinate. For example, this problem can arise
in QM/MM when MM atoms wander into the QM region.\cite{Lin2006}
In active space methods, this issue can manifest as the intruder state problem.\cite{Glaesemann1999}
Recently, in the context of active space methods, progress has been
made in automated selection of active spaces in based on a user-specified
set of relevant atomic orbitals~\cite{Sayfutyarova2017} or based
on a correlated wavefunction ansatz less expensive than the full active
space method.\cite{Stein2016,Stein2017,Dutta2017,Bao2018} These methods
have demonstrated a robust ability to select consistent active spaces
across reaction coordinates. 

In this work, we present a method for handling the subsystem inconsistency
problem in the context of wavefunction-in-DFT embedding, where a subset
localized occupied molecular orbitals (LMOs) is selected for embedded
wavefunction treatment. We begin with a previously-established charge-based
criterion for automated selection of these embedded orbitals. We next
demonstrate how \textendash{} even for a simple S\textsubscript{N}2
reaction \textendash{} this procedure can result in 
a set of embedded LMOs which is inconsistent with respect to 
the reactant and product.
Drawing inspiration
from the domain merging method for local correlation,\cite{Mata2006}
we propose an ``even-handed'' LMO selection procedure. Our method
seeks to form a consensus set of LMOs which contains for every geometry
every orbital that is important for any geometry. This results in
an automatic procedure which uses information available at the DFT
level and requires no user input beyond the set of atoms to be embedded.
Although we present this method in the context of projection-based
embedding specifically,\cite{Manby2012,Goodpaster2014a} the methodology
is sufficiently general to 
be applied in other embedding contexts.

\section{Theory\label{sec:Theory}}

\subsection{Projection-based embedding}

We begin by reviewing the projection-based embedding method, which
is a rigorous framework for embedding a correlated wavefunction theory
in a mean field theory (MF) such as Hartree Fock theory (HF) or DFT, with
interactions between the embedded and embedding densities treated
at the MF level. The system is partitioned into two subsystems: subsystem
$A$ is generally treated at the correlated wavefunction level and
subsystem $B$ treated at the MF level. Projection-based embedding
belongs to a class of methods which ensure orthogonality between the
MF description of subsystems $A$ and $B$.\cite{Manby2012,Goodpaster2014a}
This class includes frozen-core
approximations~\cite{Lykos1956}, the region method for local correlation~\cite{Mata2008},
Henderson's coupled cluster in DFT embedding method~\cite{Henderson2006},
Khait and Hoffman's modified KSCED,\cite{Khait2012}
the Huzinaga projection operator method,~\cite{Hegely2016}
and more recently the multi-level
Hartree Fock~\cite{Saether2017} and OCBSE methods~\cite{Culpitt2017}. 

As input to a projection-based calculation, we specify a low-level
MF method, a high-level correlated wavefunction method, and a set
of atoms to be embedded, $\left\{ X\right\} _{A}$. The algorithm
begins by performing a MF calculation on the full system, resulting
in a set of occupied molecular orbitals, which are then rotated to
form a set of LMOs, $\left\{ \psi\right\} $. The LMOs are partitioned
into two sets, $\left\{ \psi\right\} _{A}$ and $\left\{ \psi\right\} _{B}$,
corresponding to subsystems $A$ and $B$, respectively. This partitioning
is usually performed by choosing $\left\{ \psi\right\} _{A}$ to be
the set of LMOs with sufficient population on $\left\{ X\right\} _{A}$
(detailed in Sec. \ref{subsec:Naive-LMO-selection}). The WF calculation
is then performed with the number of electrons necessary to occupy
the $\left\{ \psi\right\} _{A}$, and a modified core Hamiltonian\cite{Manby2012,Goodpaster2014a}
written in the atomic orbital (AO) basis as 
\begin{equation}
\mathbf{h}^{A\,\mathrm{in}\,B}=\mathbf{h}+\mathbf{g}\left[\boldsymbol{\gamma}^{A}+\boldsymbol{\gamma}^{B}\right]-\mathbf{g}\left[\boldsymbol{\gamma}^{A}\right]+\mu\mathbf{P}^{B},\label{eq:hemb}
\end{equation}
where $\mathbf{{h}}$ is the core Hamiltonian of the full system,
$\boldsymbol{\gamma}^{A}$ and $\boldsymbol{\gamma}^{B}$ are the one-particle reduced density
matrices (1RDMs) corresponding to $\left\{ \psi\right\} _{A}$ and
$\left\{ \psi\right\} _{B}$, respectively, and $\mathbf{g}$ includes
all mean-field two-electron interactions. The last term enforces orthogonality
between the subsystems using a level shift, $\mu$, and a projector
onto subsystem $B$, 
\begin{equation}
\mathbf{P}^{B}=\mathbf{S}\boldsymbol{\gamma}^{B}\mathbf{S},
\end{equation}
where $\mathbf{S}$ is the AO overlap matrix. In the limit $\mu\rightarrow\infty$,
subsystems $A$ and $B$ are exactly orthogonal; in practice, this
procedure is numerically robust for $\mu$ between $10^{4}$ and $10^{7}$
hartree and $\mu=10^{6}$ hartree is usually sufficient to converge
the embedding to a type-in-type error of less than a microhartree.\cite{Manby2012,Goodpaster2014a}

\subsection{Charge method for LMO selection\label{subsec:Naive-LMO-selection}}

In this subsection, we present the current ``charge''
algorithm that has previously proven effective for partitioning the
LMOs into $\left\{ \psi\right\} _{A}$ and $\left\{ \psi\right\} _{B}$~\cite{Manby2012,Bennie2015}.
We then construct a simple example where this method breaks down,
motivating the ``even-handed'' solution presented in Sec. \ref{subsec:Even-handed-LMO-selection}. 

The chemically intuitive notion of embedded atoms must be translated
into a specific set of LMOs, which do not fully reside on specific
atoms. The charge method includes in $\left\{ \psi\right\} _{A}$
all occupied LMOs with significant population on the $\left\{ X\right\} _{A}$.
This selection is performed using a charge threshold, $q$, 
\begin{align}
\left\{ \psi\right\} _{A} & =\left\{ \psi_{i}%
|Q_{A}\left(\psi_{i}\right)>q\right\} \nonumber \\
\left\{ \psi\right\} _{B} & =\left\{ \psi_{i}%
|Q_{A}\left(\psi_{i}\right)\leq q\right\} ,\label{eq:chargepartitioning}
\end{align}
where $Q_{A}\left(\psi_{i}\right)$ is the charge of LMO $\psi_{i}$
assigned to the atoms in $\left\{ X\right\} _{A}$ and $q$ is typically
chosen to be 0.4. For example, charges may be computed using the gross
Mulliken population~\cite{Mayer2003}
\begin{equation}
Q_{A}(\psi_{i})=\sum_{X_{j}\in\left\{ X\right\} _{A}}\sum_{\lambda\,\mathrm{{on}}\,\text{ }X_{j}}\sum_{\kappa}\gamma_{\lambda\kappa}^{\psi_{i}}S_{\kappa\lambda},\label{eq:charge-subA}
\end{equation}
where $\kappa$ indexes all AOs, $\lambda\,\mathrm{{on}}\,~X_{j}$
indexes AOs $\lambda$ centered on atom $X_{j}$, $\mathbf{S}$ is
the AO overlap matrix, and $\boldsymbol{\gamma}^{\psi_{i}}$ is the 1RDM corresponding
to $\psi_{i}$. (Alternative population schemes, such as IBO,\cite{Knizia2013IBO}
can be applied similarly.) For a single-point energy calculation,
this algorithm generally selects a reasonable $\left\{ \psi\right\} _{A}$.
However, because it considers only a single geometry at a time, problems
can arise during large molecular geometry changes.

One problem is that the number of orbitals in $\left\{ \psi\right\} _{A}$ selected
by the charge method can differ at different molecular geometries.
Because of the difference in electron chemical potentials between
the high-level method of subsystem $A$ and the low-level method of
subsystem $B$, a change in number of LMOs in $\left\{ \psi\right\} _{A}$
versus $\left\{ \psi\right\} _{B}$ manifests as a discontinuity in
the potential energy. This problem is easily remedied by choosing
$\left\{ \psi\right\} _{A}$ to be the $N$ orbitals corresponding
to the $N$ largest values of $Q_{A}$ at each geometry; $N$ is chosen
to be the largest size of $\left\{ \psi\right\} _{A}$ selected by
Eq. \ref{eq:chargepartitioning} for any geometry along the reaction
coordinate. In the following, all results presented for the charge method
include this simple fix.

\begin{figure}
\includegraphics[width=1\columnwidth]{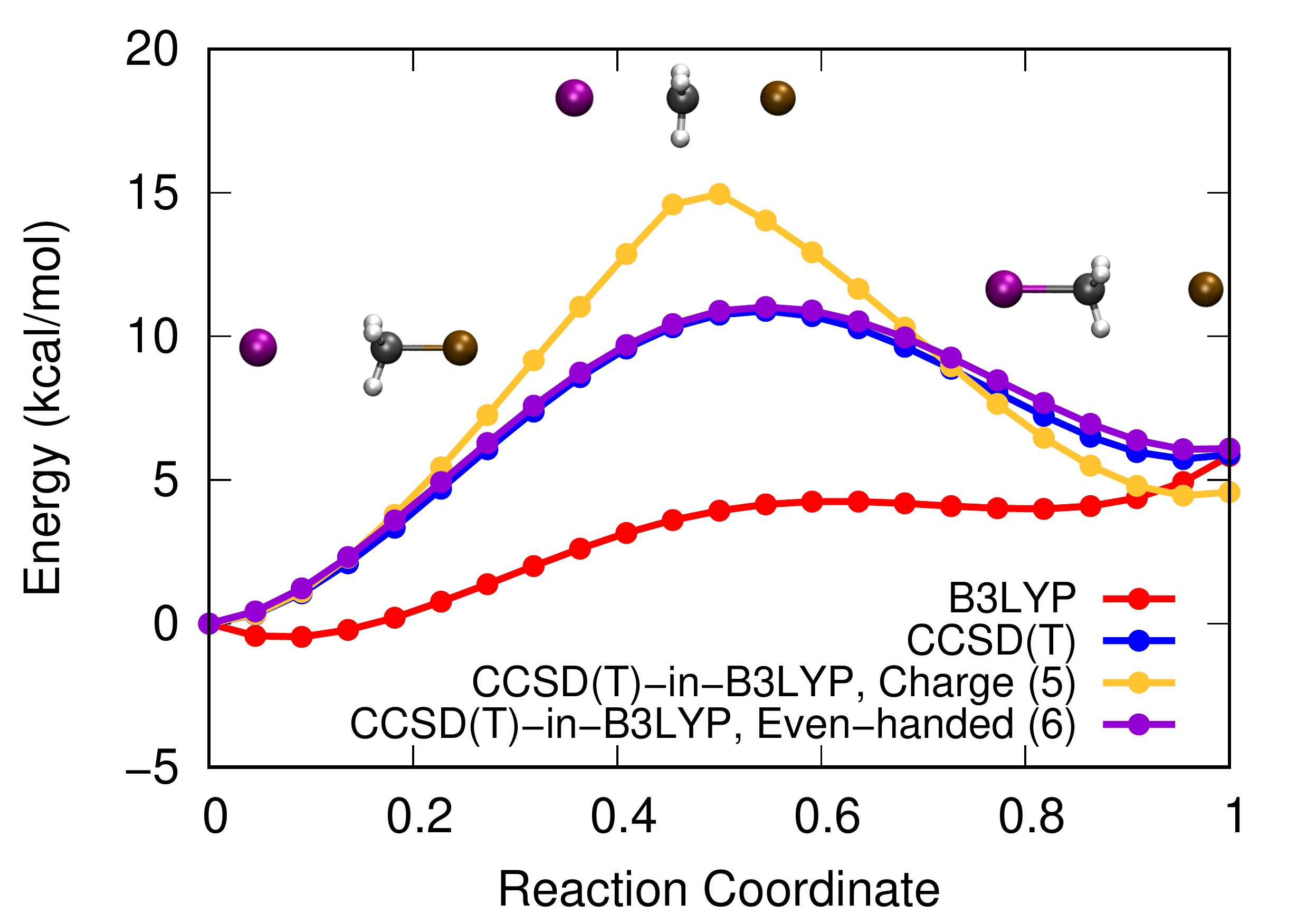}

\caption{Energy profiles for the S\protect\textsubscript{N}2 reaction of I$^-$
with bromomethane, comparing B3LYP, CCSD(T), and embedded CCSD(T)-in-B3LYP
using charge and even-handed LMO selection. The set of embedded
atoms, $\left\{ X\right\} _{A}$, contains only the carbon atom. All
curves are referenced to an energy of zero at the reactant geometry.
The number of occupied LMOs in subsystem $A$ is given in parenthesis.
\label{fig:SN2PES} }
\end{figure}

A second \textendash{} and less trivial \textendash{} problem occurs
when LMO populations on the embedded atoms change qualitatively between
geometries. Consider for example the S\textsubscript{N}2 reaction
of I$^-$ with bromomethane with $\left\{ X\right\} _{A}$
chosen to be the carbon atom. Fig. \ref{fig:SN2PES} shows the reaction
profile for coupled cluster singles and doubles with perturbative
triples, CCSD(T),\cite{Bartlett1990} embedded in the B3LYP density
functional\cite{Vosko1980,Lee1988,Becke1993,Stephens1994}, using
LMOs selected by the charge method. The result of this selection can
be seen in the embedding potential energy surface for this reaction.
Near either the reactant or the product, the CCSD(T)-in-B3LYP energy
profile follows the CCSD(T) result. However, as the transition state
is crossed, an unphysically high barrier appears in the energy profile.

\begin{figure}
\includegraphics[width=1\columnwidth]{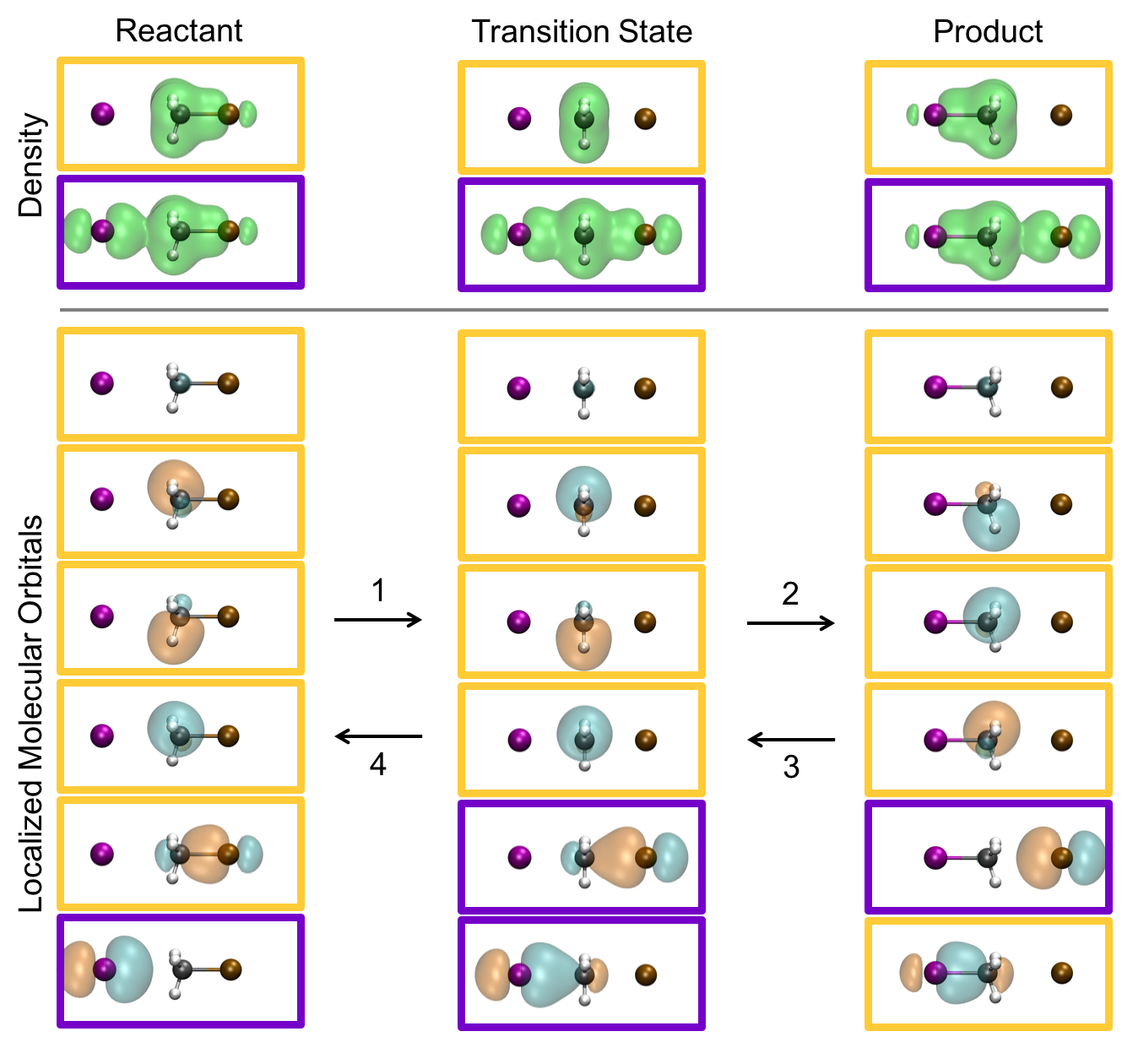}

\caption{Illustration of the even-handed selection method applied to the S\protect\textsubscript{N}2
reaction of I\protect\textsuperscript{-} with bromomethane, where
$\left\{ X\right\} _{A}$ is the carbon atom. 
Above: the subsystem $A$ densities, $\rho^A$, resulting from the charge (even-handed) method
are boxed in yellow (purple). 
Below: the set of LMOs selected for inclusion
in subsystem $A$ , $\left\{ \psi\right\} _{A}$, which comprise the corresponding $\rho^A$ above. 
In the reactant state, the charge method
$\left\{ \psi\right\} _{A}$ includes the LMO corresponding to the
C-Br bond, but not that corresponding to the C-I bond. In the product
state, the latter is included in $\left\{ \psi\right\} _{A}$ while
the former is excluded. Near the transition state, neither orbital
is included. Sweeping from reactant to product and back and taking
the union of charge method LMOs between neighboring geometries
results in the even-handed set of subsystem $A$ orbitals.\label{fig:SN2orbs}}
\end{figure}

Figure \ref{fig:SN2orbs} shows the embedded density, $\rho_A$, that results from the
charge selection method (top panel, highlighted in yellow). 
At geometries near the reactant, $\rho_A$ resembles the density of bromomethane;
near the product, $\rho_A$ resembles iodomethane. 
In the transition region, $\rho_A$ switches abruptly between these two qualitatively different densities.
Examining the orbitals $\left\{ \psi\right\}_{A}$ that comprise $\rho_A$ at each geometry (bottom panel, yellow), we see that the reactant, transition state, and product differ
on whether the LMOs associated with the C-Br and C-I $\sigma$ bonds are included in $\left\{ \psi\right\}_{A}$, with reactant-like geometries including the C-Br bond LMO and excluding the C-I bond LMO, and vice versa for product-like geometries. 
In the transition region, the abrupt change in orbital character from C-Br to C-I  
results in an incorrectly high reaction barrier. 
Although this problem is presented in the context
of projection-based embedding, we note that it is common to any embedding
method which relies on atomic populations to partition orbitals. 

\subsection{Even-handed LMO selection\label{subsec:Even-handed-LMO-selection}}

This problem can be solved by ensuring that $\left\{ \psi\right\}_{A}$ 
spans a consistent volume of Hilbert space as a function of geometry,
and thus provides a  
consistent description of $\gamma_A$ throughout a reaction profile. 
The solution requires specification of 
(i) a strategy for determining what the $\left\{ \psi\right\}_{A}$ should be 
and 
(ii) a method for quantitatively comparing $\left\{ \psi\right\}_{A}$ between geometries. 

To address the first point, we propose an ``even-handed'' procedure. Starting
with a reaction coordinate composed of an ordered set of geometries,
the charge selection procedure is performed at every geometry. We then
form the union of these charge-selected orbitals. To do this, we
sweep from reactant to product and back. For each geometry $k$ along this
sweep, the corresponding $\left\{ \psi\right\}^k_{A}$ is augmented
to include any orbitals that both match a member of the $\left\{ \psi\right\}^{k-1}_{A}$
corresponding to the previous geometry (on the basis of the method for quantitatively comparing orbitals between geometries, described later) and that are not already
included in $\left\{ \psi\right\}^{k-1}_{A}$. 

Figure \ref{fig:SN2orbs} illustrates the application of this procedure
to the aforementioned example S\textsubscript{N}2 reaction. For visual
simplicity, we consider three geometries: the reactant $R$, the transition
state $T$, and the product $P$. Four charge-selected LMOs are common
to $\left\{ \psi\right\} _{A}^{R}$, $\left\{ \psi\right\} _{A}^{T}$,
and $\left\{ \psi\right\} _{A}^{P}$. Sweeping along the reaction
coordinate from $R$ to $P$, we begin by comparing $\left\{ \psi\right\} _{A}^{R}$
to $\left\{ \psi\right\} _{A}^{T}$. $\left\{ \psi\right\} _{A}^{R}$
contains a C-Br \textgreek{sv} bond LMO that $\left\{ \psi\right\} _{A}^{T}$
lacks; the matching LMO in $\left\{ \psi\right\} ^{T}$
is thus added to $\left\{ \psi\right\} _{A}^{T}$, resulting in a five orbital
set (labeled in the figure as step 1). Comparing this updated $\left\{ \psi\right\} _{A}^{T}$
to $\left\{ \psi\right\} _{A}^{P}$, we see that $\left\{ \psi\right\} _{A}^{P}$
lacks a C-Br \textgreek{sv} bond LMO and contains an additional C-I
\textgreek{sv} bond LMO absent in $\left\{ \psi\right\} ^{T}$; the
matching C-Br \textgreek{sv} bond LMO is thus added to $\left\{ \psi\right\} _{A}^{P}$,
resulting in a six orbital set (labeled step 2). We now sweep backwards from
$P$ to $R$. $\left\{ \psi\right\} _{A}^{P}$ is compared $\left\{ \psi\right\} _{A}^{T}$,
and the latter is found to lack the C-I \textgreek{sv} bond LMO, which
is added to form a new $\left\{ \psi\right\} _{A}^{T}$ (labeled step 3).
Finally, $\left\{ \psi\right\} _{A}^{T}$ is compared $\left\{ \psi\right\} _{A}^{R}$
and the C-I \textgreek{sv} bond LMO is added to the latter (labeled step 4).
The result is that for all three geometries, $\left\{ \psi\right\} _{A}$
contains the four methyl LMOs, the C-Br \textgreek{sv} bond LMO, and
the C-I \textgreek{sv} bond LMO. 
Examining the $\rho_A$ resulting from even-handed LMO selection (Figure \ref{fig:SN2orbs}, top panel, purple), we see that it %
remains consistent from reactant to transition state to product, especially in comparison to the $\rho_A$ obtained from the charge selection method (yellow).
Figure \ref{fig:SN2PES} shows that
the even-handed scheme corrects the error in the CCSD(T)-in-B3LYP
energy profile generated using the charge selection method, yielding a CCSD(T)-in-B3LYP
energy profile in quantitative agreement with a CCSD(T) calculation
on the full system. In the following section, we will examine the relative performance of even-handed selection versus charge selection with regard to a given increase in the size of  $\left\{ \psi\right\} _{A}$.

To complete the even-handed LMO selection algorithm, 
we must define a way to quantitatively compare LMOs 
$\left\{ \psi\right\}_{A}^{k}$ and $\left\{ \psi\right\}_{A}^{k+1}$ associated with neighboring geometries along the reaction coordinate. 
The LMOs for
geometry $k$ do not directly correspond to those of $k+1$; they
have different basis functions (due to
motion of the atom-centered gaussians) and different LMO coefficients. The
latter problem is addressed with a maximum overlap formalism, similar
to that used in non-Aufbau MF calculations.~\cite{Gilbert2008} Specifically,
for geometry $k+1$ we form a new subset of $\left\{ \psi\right\} ^{_{k+1}}$
which contains the $M$ orbitals that best overlap with the span of
the $M$ orbitals so far included in subsystem $A$ for geometry $k$,
$\left\{ \psi\right\} _{A}^{k}$. For each LMO $\psi_{i}\in\left\{ \psi\right\} ^{k+1}$,
we assign an overlap with the span of $\left\{ \psi\right\} _{A}^{k}$, 

\begin{equation}
o_{i}^{k+1}=\sum_{j}^{N_{occ}}\left(\sum_{\lambda\kappa}^{N_{AO}}L_{\lambda i}^{k}S_{\lambda\kappa}^{k,k+1}L_{\kappa j}^{k+1}\right)^{2},\label{eq:MOM}
\end{equation}
where $N_{AO}$ is the number of basis functions, $N_{occ}$ is the
number of occupied orbitals, $\mathbf{L}^{k}$ is the LMO coefficient
matrix corresponding to geometry $k$, and $\mathbf{S}^{k,k+1}$ is
an overlap matrix between the AO bases corresponding to the two geometries. 

Because the two geometries do not share the same AO basis, %
the precise definition of
$\mathbf{S}^{k,k+1}$ is not immediately obvious. We might choose
the spatial overlap of the AOs; however, due to the exponential tails
of gaussian basis functions, geometries must be very closely spaced
to ensure numerical stability. Instead, we compare the orbitals between
two different geometries as though the basis had not moved. In order
to correct for changes in AO overlap, we introduce L\"{o}wdin orthogonalization,
resulting in

\begin{equation}
\mathbf{S}^{k,k+1}=\left(\mathbf{S}^{k}\right)^{\frac{1}{2}}\left(\mathbf{S}^{k+1}\right)^{\frac{1}{2}},\label{eq:overlap}
\end{equation}
where $\mathbf{S}^{k}$ and $\mathbf{S}^{k+1}$ are the AO overlap
matrices for geometries $k$ and $k+1$. As we sweep back and forth
across the reaction coordinate, at each geometry $k+1$ we form a
new set of LMOs $\left\{ \psi\right\} _{A}^{k+1}$ which contains
the $M$ LMOs in $\left\{ \psi\right\} ^{k+1}$ corresponding to the
$M$ largest $o_{i}$ as well as any charge-selected orbitals missed
by this criterion. The even-handed selection algorithm is summarized
in Algorithm \ref{alg:even-handed}.

\begin{algorithm}[H]     
\caption{Even-handed LMO selection.}     
\label{alg:even-handed}
\begin{algorithmic}[1]
\State{\textbf{Input}: LMOs $\left\{\psi\right\}^k$ for each geometry $k$, $k=1,...,N$.}
\State{For each geometry, select $\left\{\psi\right\}^k_A$ using the charge method (Eq. \ref{eq:chargepartitioning}).}
\For{$k=1,...,N-1$} 
\State{Compute $o_i^{k+1}$ using Eq. \ref{eq:MOM}.}
\State{Let $M$ be the number of LMOs in $\left\{\psi\right\}^{k}_A$.}
\parState{Construct $\left\{\psi\right\}^{k+1}_\mathrm{EH}$ as the $M$ LMOs in $\left\{\psi\right\}^{k+1}$ \\corresponding to the $M$ largest $o_i^{k+1}$.}
\State{Set $\left\{\psi\right\}^{k+1}_{A}$ = $\left\{\psi\right\}^{k+1}_A \cup \left\{\psi\right\}^{k+1}_\mathrm{EH}$}
\EndFor
\State{Repeat the above ``for loop'' %
with the order of the geometries reversed.}
\State{Repeat the forward and reverse ``for loops'' until the number of selected LMOs converges, typically after one cycle.}
\end{algorithmic}
\end{algorithm}

A potential concern is that the overlap of Eq. \ref{eq:overlap} could
become inaccurate if neighboring geometries are spatially distant.
A useful diagnostic of whether two neighboring geometries are too
distant can be formulated by comparing the $M$th largest $o_{i}$
to the ($M+1$)th largest $o_{i}$ of Eq. \ref{eq:MOM} after the even-handed
procedure is complete. This corresponds to the difference in overlap
between the least overlapping LMO included in $\left\{ \psi\right\} _{A}$
and most overlapping LMO excluded from $\left\{ \psi\right\} _{A}$.
If this difference becomes small, additional geometry points may
be added along the reaction coordinate. These interpolating geometries require only MF calculations
and thus have negligible impact on the overall cost
of the calculation. For the systems studied in this work, this problem
does not arise; 
for the energy profiles presented, 
we choose the density of interpolating geometries to illustrate the smoothness
of energy profiles rather than to minimize the number of interpolating geometries.

It is worth emphasizing that the even-handed LMO selection procedure 
only uses information from MF calculations and only involves computation of
orbital overlaps; 
even-handed LMO selection thus has a negligible effect on the overall cost of 
a projection-based wavefunction-in-DFT embedding calculation. 
It is also worth noting that even-handed LMO selection introduces no additional parameters beyond that appearing in the original charge method for LMO selection and the choice of the reaction pathway.

\subsection{Even-handed AO truncation \label{subsec:aotrunc}}

Up to this point, we have provided a framework for reducing the number
of occupied orbitals in the embedded calculation, while retaining
all virtual orbitals. However, correlated wavefunction calculations,
usually scale more strongly with the number of virtual orbitals than
occupied. Noting that the number of virtual orbitals increases with
the size of the AO basis set, a useful approach is to remove the AOs
least necessary to represent $\boldsymbol{\gamma}^{A}$~\cite{Barnes2013,Bennie2015}.
A charge-based heuristic, similar in spirit to Eq. \ref{eq:charge-subA},
is used to select which AOs are necessary to represent $\boldsymbol{\gamma}^{A}$
and therefore retained. And similar to charge selection above,
this heuristic can lead to different basis sets for different geometries.
To correct this error, for all geometries we retain the union of all
AOs heuristically retained at each geometry. That is, at every geometry
we use the same truncated basis set, $\left\{ \lambda\right\} $, 

\begin{equation}
\left\{ \lambda\right\} =\bigcup_{k}\left\{ \lambda\right\} ^{k},
\end{equation}
where $\left\{ \lambda\right\} ^{k}$ is the set of AOs retained at
geometry $k$. In the systems presented in this work which contain up to 1772 basis functions, 
this procedure results in retention of no more than 20 additional basis functions compared to
standard AO truncation.~\cite{Bennie2015}

\FloatBarrier
\section{Results and Discussion}

\subsection{Intramolecular hydrogen bonding in a CO\protect\textsubscript{2}
reduction complex\label{subsec:Intramolecular-hydrogen-bonding}}

\begin{figure}
\includegraphics[width=1\columnwidth]{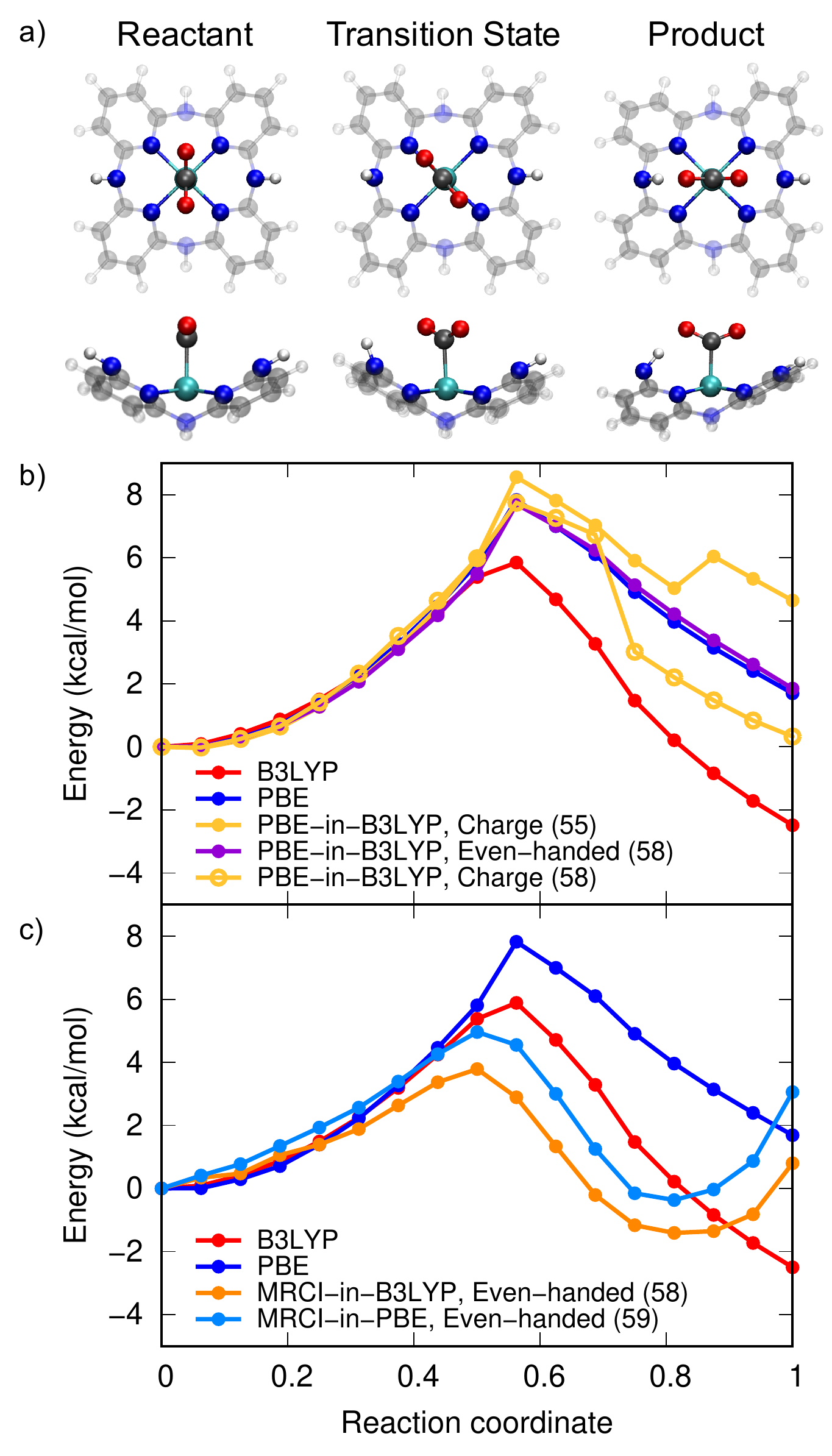}

\caption{Embedding calculations on the formation of a hydrogen bond between
a cobalt aminopyridine complex and a bound CO\protect\textsubscript{2}
molecule. (a) Reactant, transition state, and product geometries illustrated
with opaque subsystem $A$ atoms and transparent subsystem $B$ atoms.
Two views are shown for each geometry. (b) B3LYP, PBE, and PBE-in-B3LYP
energy profiles, with three methods of LMO selection for the last.
(c) Energy profiles from MRCI embedded in B3LYP and in PBE using even-handed
LMO selection. All curves are referenced to an energy of zero at the
reactant geometry. The number of occupied LMOs in subsystem $A$ is given
in parenthesis (out of 121 total). 
\label{fig:smaranda}}
\end{figure}

As a first example of even-handed partitioning, we consider a cobalt
aminopyridine complex that has been shown to catalyze the two-electron
two-proton reduction of carbon dioxide to carbon monoxide\cite{Chapovetsky2016,Chapovetsky2018}.
Experimental and DFT studies on analogous Co, Fe, and Ni complexes
suggest that CO\textsubscript{2} is stabilized by intramolecular
hydrogen bonds from pendant protons, and that these protons may transfer
to the CO\textsubscript{2} during the reduction process.\cite{Fujita1993,Froehlich2012,Costentin2014}
However, crystallography on this particular aminopyridine complex
suggests that, prior to CO\textsubscript{2} binding, the pendant protons
point away from the binding pocket, necessitating a conformational rearrangement
in order for intramolecular hydrogen bonds to form. 
The energetics of this rearrangement provide mechanistic insight into the
catalytic role of hydrogen-bonding and intramolecular proton transfer in this system. 
However, the energetics of this rearrangement presents challenges for DFT;
although widely used for Co-based systems,\cite{Leung2010,Nielsen2010,Liu2012,Peterson2012,Cheng2013,Tripkovic2013,Zhang2014,Cheng2015,Gottle2016,Shen2016}
DFT can be inaccurate for reaction barriers, 
particularly those involving multireference character. \cite{Barden2000,Diedrich2005,Jensen2007,Goel2008,Cramer2009}
Further, the geometry and strength of hydrogen bonds can be strongly
functional dependent~\cite{Santra2007,DiLabio2013}. 

Here, we employ projection-based wavefunction-in-DFT embedding to better
capture these physical effects with a correlated wavefunction. 
This approach has previously been used to study a similar Co-centered
hydrogen evolution catalyst.\cite{Huo2016} We specifically consider
the conformational change associated with forming one intramolecular
hydrogen bond. This rearrangement requires umbrella flipping of a
pendant nitrogen center and rotation of the CO\textsubscript{2} ligand
about the dihedral axis of the Co-C bond. A guess minimum energy path
is computed at the B3LYP/6-31+G{*}~\cite{Rassolov2001} level with
a Lebedev (75,302) exchange-correlation grid~\cite{Lebedev1977}
using the freezing string method~\cite{Behn2011} as implemented
in the \textsc{QChem} 4.1 software package\cite{Shao2015} 
with 15 nodes and 3 gradient descent steps. 
The system is a doublet and treated with 
an unrestricted reference. 
The path is refined by relaxing each image independently,
holding fixed the hydrogen bond length and the Co-C bond dihedral. 

The resulting B3LYP energy profile is shown in Fig. \ref{fig:smaranda}b,
with stationary-point geometries illustrated (Fig. \ref{fig:smaranda}a). 
These profiles are computed with a restricted open-shell reference for comparison with embedding calculations (below); 
as a result, the B3LYP profile does not completely reach the product minimum.
For comparison, a second profile is shown corresponding to PBE~\cite{Perdew1996a}
energy calculations performed at the same geometries. Interestingly,
the barrier predicted by PBE is larger than that predicted by B3LYP,
running contrary to the general trend that hybrid functionals predict
larger barriers than generalized gradient approximation functionals
(GGAs),\cite{Mahler2018} due to the latter over-stabilizing the relatively delocalized
electronic structure of the transition state;\cite{Patchkovskii2002}
consistently, HF predicts a still lower barrier of 3.5 kcal/mol. In this
system, the reactant and product feature electronic structures with
a greater degree of delocalization than in that of the transition
state. Thus, the self-interaction error present in GGAs may be over-stabilizing
the reactant and product relative to the transition state, resulting
in a barrier that is too high. The exact exchange present in B3LYP partially
corrects this error, reducing the barrier. 

With this reaction coordinate in hand, the even-handed procedure is used to select $\left\{ \psi\right\} _{A}$
corresponding to the choice of $\left\{ X\right\} _{A}$ shown as
opaque atoms in Fig. \ref{fig:smaranda}a.  Embedding calculations are performed in the \textsc{Molpro}
2018.0 software package\cite{Werner2012,Molpro2018.0} with the def2-TZVP basis set, an exchange-correlation
grid threshold of $10^{-10}$, and with density fitting for both Coulomb
and exchange integrals~\cite{Polly2004} evaluated with the def2-TZVPP/JKFIT basis.\cite{Weigend2007} A restricted open-shell reference is employed for all embedding calculations. The intrinsic bond orbital
(IBO) procedure~\cite{Knizia2013IBO} is employed for orbital localization. 

We first embed PBE in B3LYP
to test the smoothness of the embedded reaction coordinate and to
test convergence of the embedded energy with respect to the size of
subsystem $A$; a correlated wavefunction calculation on this large
system would prove computationally infeasible, and so PBE is used
as a proxy.
Figure \ref{fig:smaranda}b presents results for PBE-in-B3LYP embedding
for various LMO selection procedures. Selection of LMOs using the charge method  (Sec. \ref{subsec:Naive-LMO-selection})
results in a subsystem $A$ containing 55 occupied LMOs (of 121 total)
and a discontinuous PBE-in-B3LYP energy profile (yellow circles). Even-handed LMO selection
(Sec. \ref{subsec:Even-handed-LMO-selection}) yields a continuous
PBE-in-B3LYP energy profile in quantitative agreement with the whole-system
PBE energy profile (purple circles). Even-handed selection adds only 3 LMOs to $\left\{ \psi\right\} _{A}$ 
beyond those selected by the charge method, for a total of 58. 
As a further comparison, the charge method is performed 
with a modified threshold in Eq. \ref{eq:chargepartitioning}
chosen such that every geometry has 58 selected LMOs in $\left\{ \psi\right\} _{A}$.
As for the charge method with the default threshold (0.4 electrons), this modified treatment of the charge threshold also
results in a discontinuous energy profile (yellow open circles). Taken together, these results
demonstrate that 
(i) even-handed LMO selection successfully improves
upon the charge method to yield a continuous energy profile; 
(ii) the results
of the even-handed LMO selection procedure cannot be replicated by
simply performing the charge method with a different threshold; and
(iii) the resulting embedded energy profile is in qualitative agreement with 
PBE calculations performed on the whole system, 
suggesting that the embedding calculation
is converged with respect to the choice of $\left\{ X\right\} _{A}$
and that the even-handed $\left\{ \psi\right\} _{A}$ is an appropriate
set of LMOs for embedding a correlated wavefunction method (see below).

We next embed correlated wavefunction methods using the even-handed
set of LMOs determined above. 
To reduce the size of the virtual space, even-handed AO truncation
is employed as discussed in section \ref{subsec:aotrunc} with a charge
threshold of 0.001 electrons, resulting in the retention of 713 out
of 1102 AO basis functions in this compact complex. 
Comparison of MP2 with MP2-in-B3LYP calculations confirms that
this AO truncation alters the reaction energy by less than %
0.3 kcal/mol. Initial CCSD-in-B3LYP calculations performed
at each geometry along the reaction coordinate yield T1 diagnostics~\cite{Lee2003}
of at least 0.07; this value exceeds both the standard threshold of
0.02 and the threshold of 0.05 proposed for first-row transition
metals in Ref. \citenum{Jiang2012}, suggesting the presence of strong
multireference character that CCSD cannot reliably describe.

To treat this potential multireference character, embedded multireference configuration
interaction singles and doubles (MRCI)~\cite{Knowles1988,Werner1988} in DFT
calculations with a complete active space self-consistent field (CASSCF)~\cite{Knowles1985,Werner1985}
reference are performed. 
CASSCF calculations are performed without density fitting.
The active space is chosen to comprise nine
electrons in nine orbitals. Two factors suggest this active space
is sufficiently large. First, the unrestricted natural orbital complete
active space method of Bofill and Pulay~\cite{Bofill1989} produces
a guess active space that is at most three electrons in three orbitals
at any geometry. Second, the CAS(9,9) canonical orbital populations
are at least 1.98 for the lowest-energy active orbital and at most
0.03 for the highest-energy active orbital across all geometries.
Multireference character is confirmed by substantial deviations from
integer occupations within these active canonical orbitals. All reference
configurations with norm larger than 0.01 are included in the MRCI
dynamical correlation calculation. This parameter is sufficient to
converge the resulting energy profile and results in a maximum of
52 reference configurations. Relaxed reference Davidson inextensivity
corrections are applied.\cite{Davidson1977}

Figure \ref{fig:smaranda}c shows the results of embedding MRCI in
both B3LYP and PBE. (An even-handed $\left\{ \psi\right\} _{A}$ was
determined for the case of PBE in the same manner as for the case
of B3LYP.) The resulting reaction profiles are continuous and smooth
like those of the test PBE-in-B3LYP calculations. Under the MRCI-in-DFT
embedding, the hydrogen bond is weaker and has a longer bond length
(corresponding to a smaller optimal value of the reaction coordinate)
compared to the underlying DFT calculations. The embedded profiles
also show a further decrease in the barrier height. Comparing embedding
in B3LYP versus in PBE, both result in the same qualitative shape
of the reaction profile. Moving along the reaction coordinate, the
MRCI-in-B3LYP and MRCI-in-PBE energy profiles diverge, resulting in
a disagreement of 2 kcal/mol at the geometry corresponding to a reaction
coordinate value of 1. This disagreement is smaller than the 4 kcal/mol
disagreement between B3LYP and PBE at the same geometry, but is still
significant. This residual difference is due to different descriptions
by the two functionals of the geometric changes in subsystem $B$.
Across this reaction coordinate, significant strain forms in the ligand
backbone which lies largely in subsystem $B$. Thus, we see that embedding largely removes
the dependence of the description of subsystem $A$ on the density functional employed for subsystem $B$. 

In this example, charge selection results in a qualitatively incorrect discontinuous energy profile, while even-handed selection results in a quantitatively accurate and continuous profile. This improvement comes at the cost of merely 3 additional LMOs included in $\left\{ \psi\right\} _{A}$.

\subsection{Binding of CO to Cu clusters}
\begin{figure*}
\includegraphics[width=1\textwidth]{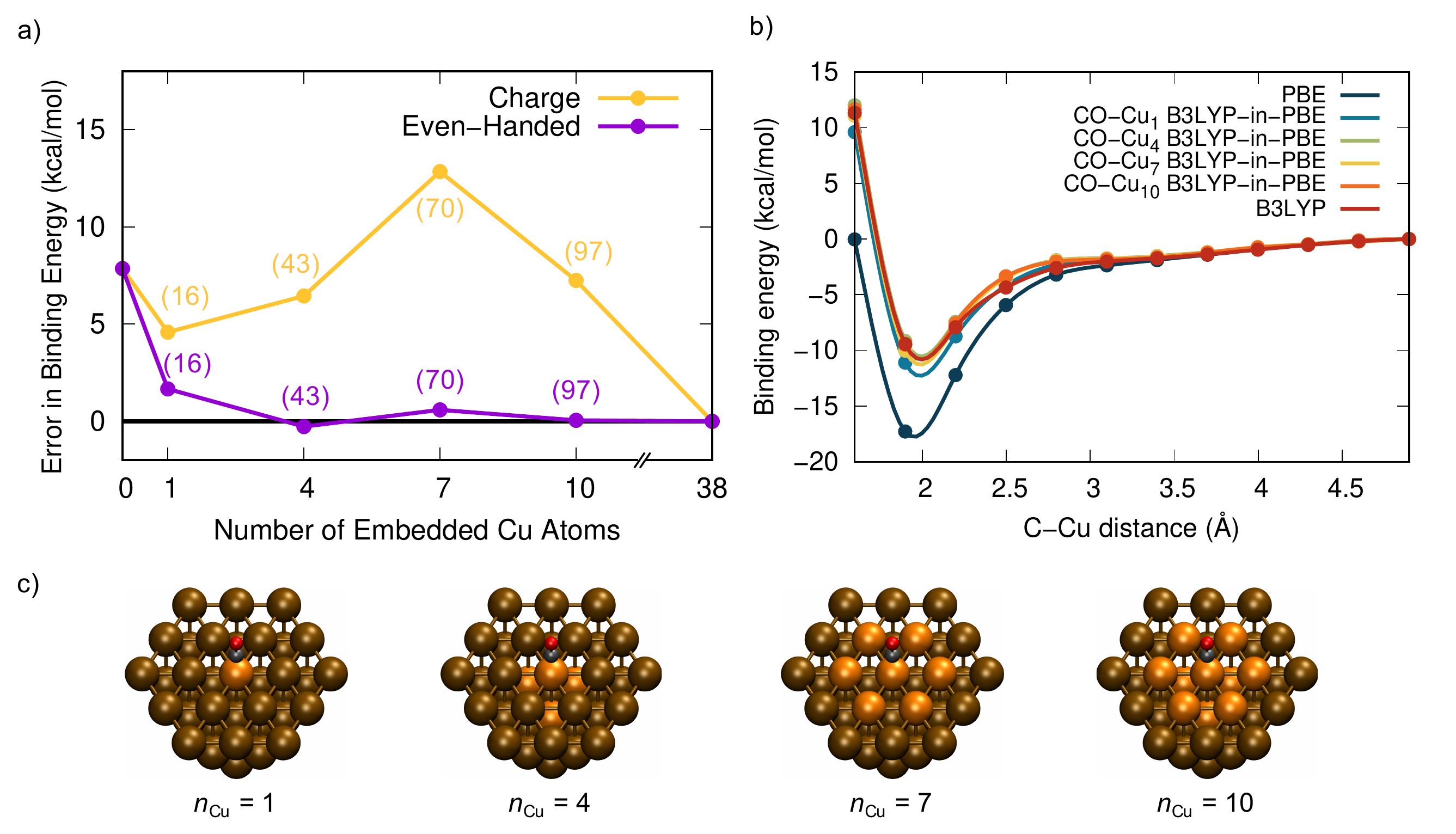}
\caption{\label{fig:Cu38}
Embedded calculations on CO-Cu\protect\textsubscript{38}
top site binding. 
(a) Convergence of the binding energy error with respect to subsystem $A$ size for B3LYP-in-PBE embedding using charge and even-handed LMO selection.
$n_{\mathrm{Cu}}=0$ corresponds to a PBE calculation on the whole system and 
$n_{\mathrm{Cu}}=38$ corresponds to a B3LYP calculation on the whole system.
The number of occupied LMOs in subsystem $A$ is given in parenthesis (out of 368 total).
As discussed is Sec. \ref{subsec:Naive-LMO-selection}, the set of LMOs selected by the charge method with a given threshold is augmented so that a consistent number of LMOs is selected across all geometries in the reaction coordinate. %
(b) CO-Cu\protect\textsubscript{38}
binding potential energy surfaces calculated using B3LYP-in-PBE embedding
with even-handed LMO selection. 
(c) CO-Cu\protect\textsubscript{38}
geometry. Subsystem $A$ contains CO as well as the Cu atoms indicated
in orange. }
\end{figure*}

As a second demonstration of even-handed selection, we consider the
binding of carbon monoxide to copper clusters. This class of systems
has proven difficult for DFT, with different functionals predicting
a wide range of binding energies as well as different preferred binding
sites. For the example of CO binding to a Cu(111) surface, PBE, which
is often accurate for metals, incorrectly predicts a hollow site preference
for CO binding~\cite{Sharifzadeh2008}. Meanwhile B3LYP, which is
often accurate for molecules, predicts the correct on-top binding
site preference, but with a binding energy of $-2.2$ kcal/mol~\cite{Neef2006},
compared to experimental measurements which range between $-10.4$ and
$-12.0$ kcal/mol~\cite{Kessler1977,Kirstein1986,Bartels1999,Vollmer2001,Patra2018}.
Previous work employing frozen-density embedding with approximate
non-additive kinetic energy potentials has shown that embedded MRCI-in-LDA
calculations can reproduce the measured CO-Cu(111) binding-site
preference and binding energy~\cite{Sharifzadeh2008}. 

We first consider the binding of CO to a Cu\textsubscript{38} cluster,
constructed as a hemispheric
cut of a Cu(111) surface centered around a top site, corresponding
to two surface neighbor rings around this site. 
A reaction path for
the binding of CO is constructed by varying the C-Cu bond length in
steps of 0.3 \AA~ and relaxing the C-O bond length at the PBE/def2-TZVP
level of theory, holding the Cu atoms fixed in their bulk lattice
positions. 
The Cu atoms are treated with the 10 electron Stuttgart/Cologne effective
core potential ECP10MDF.\cite{Figgen2005} 
As in section \ref{subsec:Intramolecular-hydrogen-bonding},
calculations are performed in the \textsc{Molpro} 2018.0 software
package with an exchange-correlation grid threshold of $10^{-10}$ and
with density fitting for Coulomb and exchange integrals evaluated with the def2-TZVPP/JKFIT basis.
The binding energy is found to be -17.3 kcal/mol with a
C-Cu bond length of 1.90 \AA. At the same geometry, the binding energy
from B3LYP is -9.4 kcal/mol.

Figure \ref{fig:Cu38} examines the convergence of the 
B3LYP-in-PBE binding energy as a function of subsystem $A$. 
Four choices of subsystem $A$ are considered, corresponding to embedding CO and 1, 4, 7, and
10 copper atoms (Fig. \ref{fig:Cu38}c).
Even-handed AO truncation is employed with a charge threshold of 0.001 electrons, 
and is found to introduce negligible errors. Figure \ref{fig:Cu38}a shows the
converge of the embedded binding energy using charge (yellow) versus even-handed LMO selection (purple). 
Charge selection results in an inconsistent set of LMOs for subsystem $A$, $\left\{ \psi\right\} _{A}$, between the
bound and unbound CO geometries, leading to large errors even for
the larger sizes of subsystem $A$ considered. By contrast, even-handed
selection results in rapid convergence to the full B3LYP result. 
Even for the smallest choice of subsystem $A$ tested ($n_{Cu}=1$), even-handed
selection reproduces the full B3LYP result within 1.5 kcal/mol, despite
treating only three atoms (8 occupied LMOs) at the B3LYP level. Notably,
the charge and even-handed methods select the same number of LMOs for
each choice of $\left\{ X\right\} _{A}$. Eq. \ref{eq:chargepartitioning}
selects the most LMOs at bound geometries, and LMOs must be added
to the $\left\{ \psi\right\} _{A}$ corresponding to unbound geometries
in order to keep the number of occupied LMOs constant throughout the binding
coordinate. Due to different polarization of the Cu atoms in the bound
versus unbound geometry, the charge method selects qualitatively
different LMOs than in the $\left\{ \psi\right\} _{A}$ corresponding
to the bound geometry. Figure \ref{fig:Cu38}b shows CO-Cu\textsubscript{38}
binding curves computed with PBE, B3LYP, and B3LYP-in-PBE embedding.
{} These B3LYP-in-PBE energy profiles are smooth and converge rapidly
to the reference B3LYP result. 
Further, these results include ligand-induced relaxation of the electronic density on the metal cluster %
that
was neglected in previous frozen-density embedding calculations.\cite{Sharifzadeh2008} 
Taken together Figs. \ref{fig:Cu38}a
and \ref{fig:Cu38}b demonstrate that even-handed LMO selection enables
rapid convergence with respect to the size of subsystem $A$ and yields
smooth energy profiles. 

B3LYP-in-PBE embedding provides some computational savings compared
to a B3LYP calculation on the whole system. Averaged over the twelve
geometries considered, B3LYP-in-PBE embedding results in speedups
\footnote{B3LYP reference timing calculations are started from both the Superposition
of Atomic Densities (SAD) guess~\cite{VanLenthe2006} and from the
converged PBE orbitals. The faster of these two timings is used for
speedup calculations. PBE and B3LYP-in-PBE timings are computed starting
from the SAD guess.} 
versus B3LYP ranging from 5.8x for $n_{Cu}=1$ to 3.4x for $n_{Cu}=10$.
(PBE is on average 7.2x faster than B3LYP for this system in the implementation used here.) 
This speedup results both from a reduction in the number of AOs for which exact
exchange is evaluated and from a reduction of the number of SCF cycles
necessary to converge the embedded B3LYP calculation.\cite{Libisch2017}

The metallic nature of this system (PBE HOMO-LUMO gap of 0.2 eV) results in poor
orbital localization, particularly for those MOs near the Fermi level with
primarily 4s character.
\footnote{The Pipek-Mezey~\cite{Pipek1989} and Boys~\cite{Boys1960} localization schemes 
result in a greater number of poorly localized LMOs, as does changing the IBO localization 
exponent from 4 to 2.} 
Consequently, AO truncation,
which depends on spatial locality of $\boldsymbol{\gamma}^A$, results in a large fraction 
of retained AO basis functions. For $n_{\mathrm{Cu}}=1$, the smallest subsystem $A$ 
considered, 789 AOs are retained (out of 1772 total);
1413 AOs are retained for the case of $n_{\mathrm{Cu}}=10$.
As a result, wavefunction-in-DFT embedding with this approach is computationally infeasible %
without further methods development regarding virtual space reduction or occupied MO localization;  
for the purposes of the current work, we instead focus on a smaller cluster with a larger band gap. 

\FloatBarrier
A Cu\textsubscript{10} cluster is constructed similarly to the Cu\textsubscript{38} cluster 
by a hemispheric cut of a Cu(111) surface centered
around a top site, corresponding to one nearest-neighbor shell around
this site. The bound CO-Cu\textsubscript{10} geometry is optimized at the PBE/def2-SVP level
of theory with the Cu atoms held fixed in their bulk lattice positions. 
The smaller size of this cluster results in an increase of the PBE HOMO-LUMO gap to 0.7 eV.

\begin{figure}
\includegraphics[width=1\columnwidth]{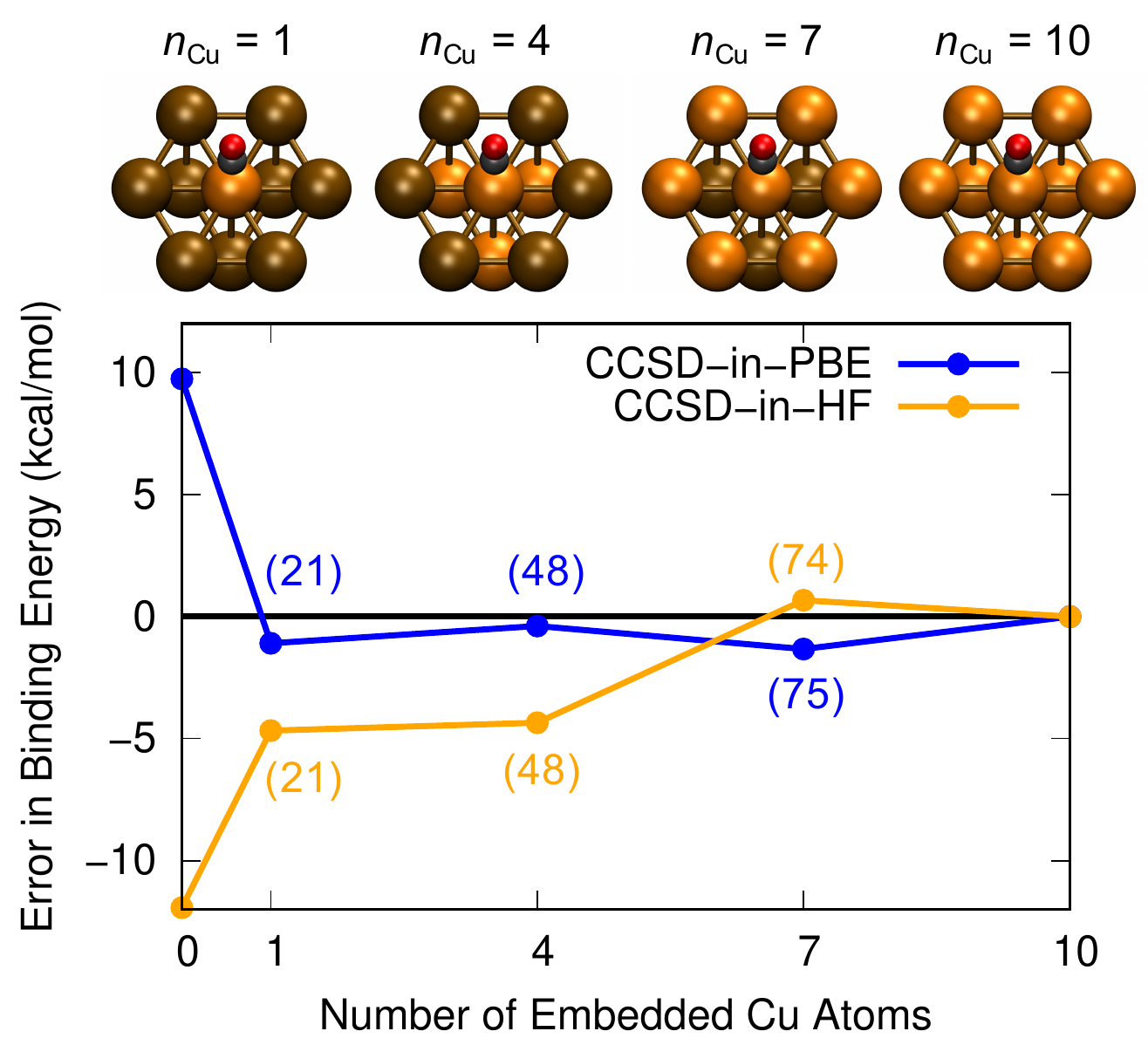}

\caption{Error in the embedded CCSD CO-Cu\protect\textsubscript{10} binding
energy with respect to subsystem $A$ size for the case of embedding
in PBE versus embedding in HF. Above: CO-Cu\protect\textsubscript{10}
geometry with subsystem $A$ Cu atoms shown in orange for the three
sizes of subsystem $A$ considered. (CO is always included in subsystem
$A$.) The number of occupied LMOs in subsystem $A$ is given in parenthesis
(out of 102 total). $n_{\mathrm{Cu}}=0$ corresponds to a MF calculation
on the whole system, while $n_{\mathrm{Cu}}=10$ corresponds to a
CCSD calculation on the whole system. \label{fig:Cu10} }
\end{figure}

CCSD-in-PBE and CCSD-in-HF embedding calculations are performed in a def2-SVP basis. 
Three choices of subsystem $A$ are constructed in the same manner as in the Cu\textsubscript{38}
cluster (see Fig. \ref{fig:Cu10}). 
$\left\{ \psi\right\} _{A}$ is chosen with even-handed LMO selection.
(Although for these three choices of $\left\{X\right\}_A$, the charge method selects the same LMOs.)
Figure \ref{fig:Cu10} presents the binding energy error associated
with CCSD-in-PBE versus CCSD-in-HF embedding as a function of the
choice of subsystem $A$. 
Errors are computed relative to reference full-system CCSD calculation, 
available due to the smaller size of this cluster.
The T1 diagnostic is found to
be less that 0.02 at all geometries. CCSD-in-PBE embedding converges
rapidly to the full CCSD result, with an error of 1 kcal/mol at the
smallest subsystem $A$ size considered. By contrast, CCSD-in-HF embedding
does not reach the same level of accuracy until subsystem $A$ contains
all atoms in the system except the three subsurface Cu atoms. As in
Fig. \ref{fig:smaranda}, projection-based embedding removes dependence
on the MF method used for subsystem $B$; however, a larger subsystem
$A$ is required when the MF poorly describes the electronic structure
of subsystem $B$ (as in HF for this Cu cluster).

Compared to the charge selection method, which is fragile to changes in polarization
upon CO binding, we see that even-handed LMO selection gives
a consistent $\left\{ \psi\right\} _{A}$ with respect to all
geometries along the binding coordinate (Fig. \ref{fig:Cu38}). 
The result is rapid convergence 
with respect to the size of subsystem $A$
of the embedded energy to the limit
of a full-system high-level calculation.

\FloatBarrier
\section{Conclusion}

Projection-based embedding provides a framework for rigorous wavefunction-in-DFT embedding. %
In this paper, we demonstrate that the standard charge-based
criterion for selecting 
the integer number of occupied LMOs that comprise the embedded subsystem
can lead to discontinuous energy
profiles and slow convergence of the calculated energy with respect
to the size of subsystem $A$. We have introduced an even-handed selection
procedure that ensures a consistent set of embedded occupied LMOs
throughout a reaction coordinate. This algorithm only uses mean-field quantities 
that have already been computed as part of the 
projection-based embedding wavefunction-in-DFT procedure; thus, it adds negligible cost.

This method has been applied in several situations including for the
cases of an organometallic catalyst and the binding of a molecule
to a metal cluster, considering both DFT-in-DFT embedding and wavefunction-in-DFT
embedding. In all cases, the even-handed  method has been shown to
be superior to the original charge method for selecting LMOs, resulting in smooth potential
embedded energy surfaces and more rapid convergence of the embedded
energy with respect to the size of subsystem $A$. Further, embedding
calculations performed using even-handed selection
are largely insensitive to the underlying mean-field theory used to treat subsystem
$B$, although it is worth noting that a sufficiently poor choice
of mean-field theory, such as HF for a metal cluster, can lead to slower convergence
with respect to the size of subsystem $A$
than when starting from a better mean-field theory (Fig. \ref{fig:Cu10}).

Projection-based embedding has recently been extended to include periodic boundary
conditions,\cite{Libisch2017,Chulhai2018} and the ideas of even-handed selection
can be naturally applied in such a framework; however, caution should
be taken in the case of wavefunction-in-DFT embedding for metals, due to slow convergence of the AO truncation in small-band-gap systems for which poorly-localized occupied LMOs contribute substantially to the correlation energy.
Regardless, we show that even-handed selection results in highly accurate 
B3LYP-in-PBE embedding for Cu binding to a Cu$_{38}$ cluster despite 
substantial changes in polarization of the metal (Fig. \ref{fig:Cu38}), and
accurate CCSD-in-PBE embedding was demonstrated for CO binding to the smaller Cu$_{10}$ cluster (Fig. \ref{fig:Cu10}).

The even-handed selection method introduced here provides a consistent
set of 
occupied  LMOs for an embedding calculation
across a reaction coordinate.
Because its input comes entirely from mean-field
calculations at each reaction coordinate geometry, the method has negligible cost compared to the original charge method for LMO selection. In cases where even-handed and charge selection yield
the same embedded occupied LMOs, the even-handed procedure provides
a measure of confidence that the embedded density is
consistent across geometries. When the two procedures differ, even-handed
selection has in all cases resulted in qualitatively superior results at the cost
of a modest number of additional embedded occupied LMOs.  %
We therefore recommend use
of the even-handed LMO selection procedure whenever considering a specified reaction pathway.

\section{Acknowledgments}

We thank Feizhi Ding and Sebastian Lee for helpful discussions. 
M.W. thanks the Resnick Sustainability Institute for a postdoctoral fellowship. 
T.F.M. acknowledges support in part from the NSF under Award CHE-1611581; additionally,
this material is based upon work performed by the Joint Center for Artificial Photosynthesis, a DOE Energy Innovation Hub, supported through the Office of Science of the U.S. Department of Energy under Award Number DE-SC0004993. 
F.R.M. is grateful for funding from EPSRC (Grant EP/M013111/1).

\bibliographystyle{apsrev4-1}
\bibliography{refs}

\end{document}